\documentstyle[aas2pp4]{article}

\lefthead{Brandner et al.}
\righthead{Ring nebula and bipolar outflows in NGC~3603}

\def\Ha{$\mbox{H}\alpha$\/}
\def\HII{H{\sc ii}}
\def\NII{N{\sc ii}}

\begin{document}

\title{Ring Nebula and Bipolar Outflows Associated with the B1.5 Supergiant 
Sher~\#25 in NGC 3603\footnote{Based on observations obtained at the European 
Southern Observatory, La Silla}}

\author{Wolfgang Brandner\altaffilmark{2}, Eva K.\ Grebel 
\altaffilmark{2,3}, You-Hua Chu\altaffilmark{3,5}, 
Kerstin Weis \altaffilmark{3,4,5}}
\affil{$^2$Astronomisches Institut der Universit\"at W\"urzburg, Am Hubland, 
D-97074 W\"urzburg, Germany}
\authoremail{brandner@astro.uni-wuerzburg.de,grebel@astro.uni-wuerzburg.de}

\affil{$^3$University of Illinois at Urbana-Champaign, Department of Astronomy,
1002 West Green Street, Urbana, IL 61801, USA}
\authoremail{chu@astro.uiuc.edu}

\affil{$^4$Institut f\"ur Theoretische Astrophysik, Tiergartenstr.\ 15, D-69121 
Heidelberg, Germany}
\authoremail{kweis@ita.uni-heidelberg.de}

\affil{$^5$Visiting astronomer, Cerro Tololo Inter-American Observatory,
NOAO, which are 
operated by AURA, Inc.,
under contract with the NSF}

\begin{abstract} 
We have identified a ring-shaped
emission-line nebula and a possible bipolar outflow centered on the B1.5
supergiant Sher \#25 in the Galactic giant 
\HII\ region NGC 3603 (distance 6 kpc). The clumpy ring around Sher \#25
appears to be tilted by 64$^\circ$ against the plane of the sky.
Its semi-major axis (position angle $\approx$ 165$^\circ$) 
is 6\farcs9 long, which corresponds to a ring diameter of 0.4 pc.
The bipolar outflow filaments, presumably located 
above and below the ring plane on either side of Sher \#25,
show a separation of $\approx$ 0.5  pc from the central star. 

High-resolution spectra show that the ring has a systemic velocity of 
V$_{\rm LSR}$ = +19 km s$^{-1}$ and a de-projected expansion velocity 
of 20 km s$^{-1}$, and that one of the bipolar filaments has an outflow 
speed of $\sim$83 km s$^{-1}$.  The spectra also show high 
[N{\sc ii}]/H$\alpha$ ratio, suggestive of strong N enrichment.  
Sher \#25 must be an evolved blue supergiant (BSG) past the red supergiant 
(RSG) stage. 
We find that the ratio of equatorial to polar mass-loss rate during
the red supergiant phase was $\approx$ 16.
We discuss the results in the framework of RSG--BSG wind evolutionary models.

We compare Sher \#25 to the progenitor of SN\,1987\,A, which it resembles 
in many aspects.

\end{abstract}

\keywords{Stars: evolution, individual (Sher \#25), mass-loss, and supergiants
                 -- supernovae: individual (SN\,1987\,A) --
                 ISM: individual (NGC 3603).
         }

\section{The Blue Supergiant Sher \#25 in NGC 3603}

Sher \#25 (Sher 1965) is a B1.5Iab supergiant (Moffat 1983) similar to 
Sk$-$69\,202, the progenitor of SN\,1987\,A.  
Sher \#25 has a visual magnitude of V $\approx$ 12\fm2--12\fm3 (e.g.,
van den Bergh 1978).  It is located at $\approx$20$''$ north of HD 97950, the 
core of the $\approx$ 4 Myr old cluster at the center of the Galactic 
giant \HII\ region NGC~3603 at a distance 
of 6--7 kpc (Clayton 1986; Melnick et al.\ 1989). 
Based on UBV CCD photometry of Sher \#25 Melnick et al.\ derived a 
visual extinction A$_V \approx 5^m$ and a distance modulus consistent 
with Sher \#25 being associated with NGC 3603.

In a recent search for emission-line objects in NGC 3603 (Brandner 
et al.\ 1997), we found a clumpy ring and a bipolar nebula around 
Sher \#25.  This ring is similar to that around SN\,1987\,A in both
size and morphology.  Follow-up spectroscopy shows N 
enrichment in the nebula around Sher \#25, suggesting that Sher 
\#25 is at a similar evolutionary stage as Sk$-$69\,202. In this 
letter, we report our observations (\S 2), discuss the physical 
structure of the nebula around Sher \#25 (\S 3) and the
evidence for Sher \#25 being associated with NGC 3603 (\S 4), and compare it 
to SN\,1987\,A (\S 5).

\section{Observations} 

\begin{figure}[ht]
\centerline{\plotfiddle{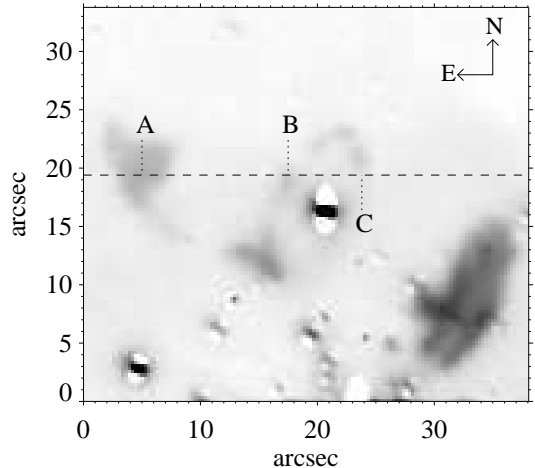}{180pt}{0}{55}{55}{-240}{0}}
\figcaption{
Continuum-subtracted \Ha\ image (NTT / EMMI,
exposure time 10 min, seeing 1$''$) of Sher \#25
 ($\alpha_{2000}$ = 11$^h$15$^m$7.8$^s$, $\delta_{2000}$ =
--61$^\circ$15$'$17$''$). The tilted ring around the blue
supergiant as well as the bipolar filaments located to the
northeast and southwest of Sher \#25 are visible.
The position of the slit is indicated by a dashed line.\label{fig2}}
\end{figure}

\begin{figure}[ht]
\centerline{\plotone{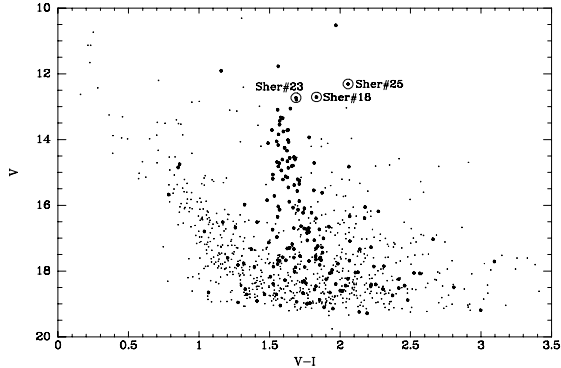}}
\figcaption{Color-magnitude diagram for NGC 3603 and
the surrounding field. Stars within the central 70$''$ $\times$ 70$''$
(2pc $\times$ 2pc) around the cluster center are marked by fat dots,
the field (foreground) star population is marked by thin dots.
The three blue supergiants are highlighted by an open ring.\label{fig3}}
\end{figure}

\Ha\ and R images of an 8$'$ $\times$ 8$'$ field centered on NGC~3603 were 
obtained at the ESO New Technology Telescope (NTT) on 1995 February 8
with the red arm of the ESO Multi-Mode Instrument and a 2k Tek CCD 
(ESO \#36). We used a narrow-band \Ha\ filter ($\Delta \lambda$=1.8 nm) 
and a broadband R filter for continuum subtraction. 
Photometric VRI observations and an H$\alpha$+[N{\sc ii}] image 
($\Delta \lambda$=6.2 nm) were obtained on 1995 March 2 with the
CCD camera at the Danish 1.54m telescope.
The H$\alpha$+[N{\sc ii}] image of the central cluster in NGC 3603 and its 
surroundings is displayed in Figure \ref{fig1}.
Figure \ref{fig2} shows a continuum-subtracted \Ha\ NTT-image
centered on Sher \#25.  The residuals (bright features) are due to
charge bleeding as the brightest stars were saturated in the R image.
Figure \ref{fig3} shows the location of the blue supergiant Sher \#25 
above and to the red of the main sequence turn-off. The stars well above the
main-sequence are (blended) Wolf-Rayet and early type O stars located
in the very center of the cluster. 

%
%

%
%

\begin{figure}[ht]
\centerline{\plotfiddle{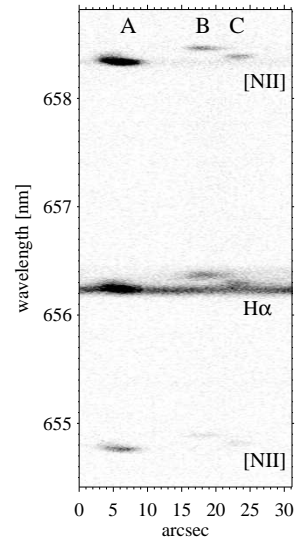}{200pt}{0}{60}{60}{-180}{0}}
\figcaption{2D high-resolution spectra of stellar
ejecta associated with Sher \#25. Two distinct velocity components
where the slit intercepts the ring (B\&C) are visible in \Ha\ and [\NII].
\label{fig4}}
\end{figure}

Long-slit echelle spectra of the eastern cap and the ring
were obtained on 1996 January 10 at the CTIO 4m telescope.
The echelle spectrograph was equipped with a 2k Tek CCD.
The slit orientation was east-west (dashed
line in Figure \ref{fig2}) and the slit width was 1\farcs6. 
We adopted rest wavelengths in air of 654.81 nm and 658.36 nm
for the two forbidden [\NII] lines $^3$P$_1$--$^1$D$_2$ and
$^3$P$_2$--$^1$D$_2$ (e.g., Moore 1959). 
Relative velocities are accurate to about 0.2 km~s$^{-1}$.
The absolute calibration with respect to the local standard
of rest (LSR) has an uncertainty of about 2--4 km~s$^{-1}$. 
Figure \ref{fig4} shows the 2D spectra in the region of the \Ha\ 
and [\NII] lines. At the position where the slit intersects the 
ring the two distinct velocity components are clearly visible.

%
%

%
%
\begin{figure}[ht]
\centerline{\plotfiddle{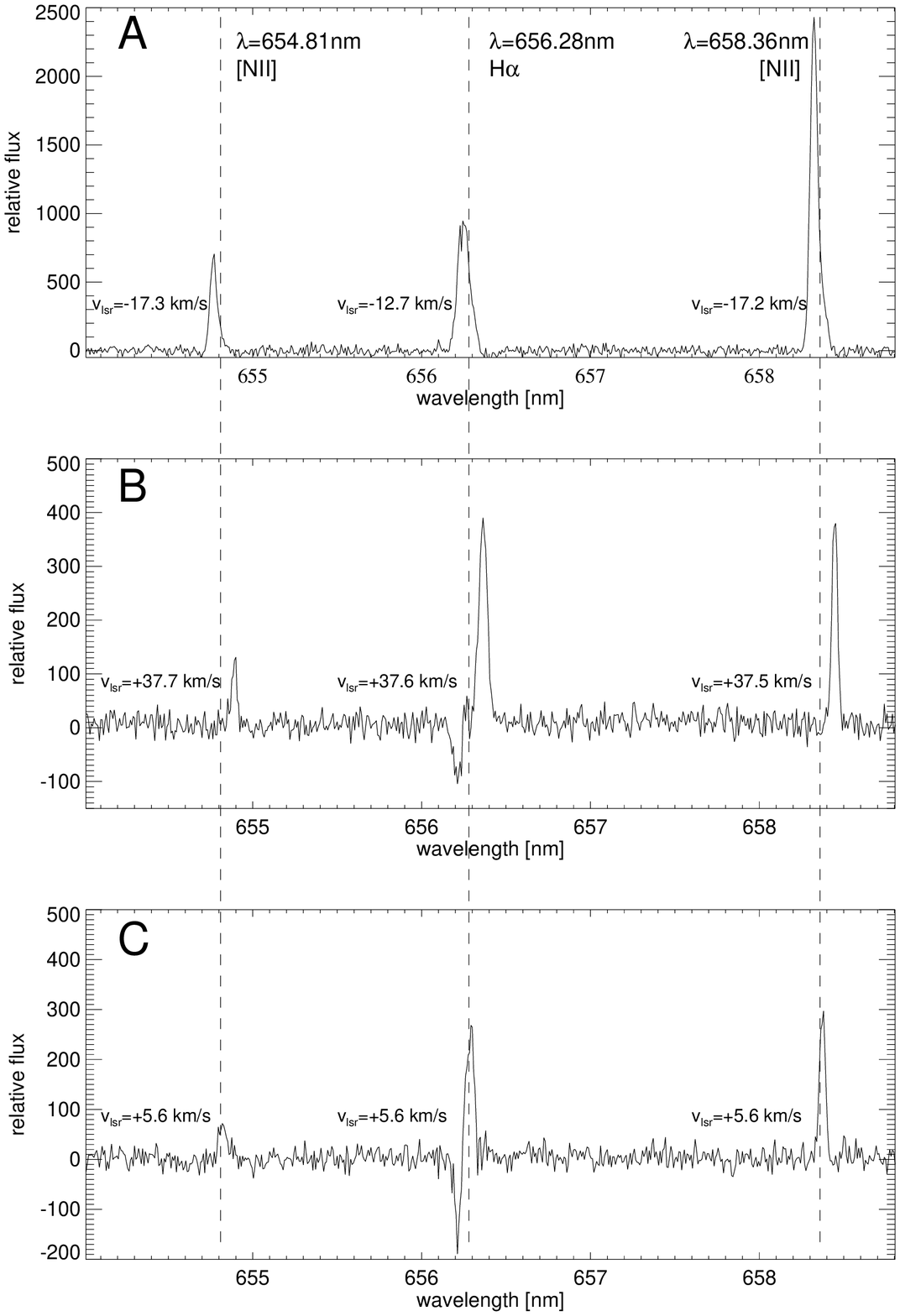}{290pt}{0}{45}{45}{-230}{0}}
\figcaption{Radial velocities of \Ha\ and [\NII] emission lines
in the outflow filament A and the two ring fragments B and C with respect to
the local standard of rest.
The dashed lines indicate the rest wavelengths of the emission lines.
The spectra are background subtracted by linear interpolation between the
nebular emission east and west of the individual knots. The absorption feature
on the blue side of the H$\alpha$ line at position B and C is due to
imperfect background subtraction.
\label{fig5}}
\end{figure}

\section{Inner Ring and Outflow Filaments around Sher \#25}

A tilted ring around Sher \#25 can be clearly seen in Figures \ref{fig1} 
and \ref{fig2}.
The semi-major and semi-minor axes are 6\farcs9 and 3\farcs05,
respectively.  Assuming a circular ring geometry, we derive an 
inclination angle of 64$^\circ$ with respect to the sky plane
along the position angle $\approx$ 165$^\circ$.
The linear diameter of the ring is 0.4 pc, for a distance of 6 kpc.
The relative velocity difference between the two ring components
intersected by the slit is 32 km~s$^{-1}$ (cf.\ Figure \ref{fig5}). Taking projection 
effects into account, we derive a de-projected expansion velocity of 20  
km~s$^{-1}$ for the ring and a systemic velocity of 
v$_{\rm LSR}$ = +19 km~s$^{-1}$.  This value agrees well with
radial velocities of molecular cloud cores south of NGC 3603
(+12 km~s$^{-1}$ to +16 km~s$^{-1}$, N\"urnberger, priv.\ comm.)
and supports Sher \#25's association with NGC 3603.

Bipolar filaments to the northeast and to the southwest of Sher \#25
are also clearly visible in Figures \ref{fig1} and \ref{fig2}. 
The northeast filament is not resolved into substructures.  It
is blue-shifted by 36.2 km~s$^{-1}$ from the systemic velocity, 
indicating an outflow nature.  
The southwestern filament shows a complex structure with two apparent
shock fronts.  Lacking kinematic information, we cannot determine 
whether the southwestern filament is part of a larger 3D structure, 
i.e., something hourglass-like.
Nevertheless, it is very likely that these bipolar filaments are 
physically produced by a bipolar outflow.
If the bipolar filaments of Sher \#25 are located along an axis 
perpendicular to the plane defined by the ring, their physical 
separation from Sher \#25 is about 0.5 pc (15$''$/$\cos 26^\circ$ 
at 6 kpc). The de-projected expansion velocity of the northeastern 
filament is $\approx$ 83 km~s$^{-1}$.

The echelle spectra of the ring and the northeast outflow filament show a high
[N{\sc ii}]/\Ha\ ratio, compared to the background H{\sc ii} region (see
Table 1 and Figure \ref{fig5}). Low-resolution spectra of the north-eastern outflow
filament indicate T$_{\rm eff} \approx$ 7000$\pm$1000 K and
$\log {\rm N}_{\rm e} \approx$ 9.8$\pm$0.3 m$^{-3}$. In a [O{\sc iii}]/H$\beta$
vs.\ [N{\sc ii}]/H$\alpha$ diagram, the north-eastern outflow filament is
situated clearly outside the location of H{\sc ii} regions or supernova
remnants (Brandner et al.\ 1997). Thus, the high [N{\sc ii}]/\Ha\ ratio is
caused by an enhanced N abundance, indicating that at least the bipolar
filaments around Sher \#25 consist of stellar material enriched by the
CNO cycle. Therefore, Sher \#25 very likely is an evolved post-red supergiant.

Given the evolutionary stage of Sher \#25, the surrounding nebula 
may be explained in the framework of interaction between red supergiant 
(RSG) wind and blue supergiant (BSG) wind.  A 2D ring-like structure 
(as opposed to 3D shells) can be produced if the density of the RSG wind 
is a strong function of polar angle, peaking along the equatorial plane 
and decreasing toward the poles (Blondin \& Lundqvist 1993; Martin \& 
Arnett 1995).
A ring develops as the fast BSG wind sweeps up the dense RSG wind material.
At the same time the density gradient in the RSG wind allows the fast BSG
wind to expand more easily in polar directions.
This process might lead to an hourglass-shaped emission nebula as has been
observed in the young planetary nebula MyCn18 (Sahai et al.\ 1995).
The clumpy structure of the ring (see Figure \ref{fig2}) very likely originates 
in Rayleigh-Taylor instabilities at the interface between the swept-up 
slow RSG wind and the fast BSG wind.

The expansion velocities and physical dimensions yield a dynamical
age of $\approx$ 9000 yr for the ring and $\approx$ 6000 yr for the
polar outflows.
Applying a self-similarity solution for the interaction regions of
colliding winds (e.g., Chevalier \& Imamura 1983) we can carry out a crude
analysis of the observed velocities. In the following we assume
constant mass loss rates and constant wind velocities for the slow wind and 
the fast 
wind, a stalled shock (pressure equilibrium) in a spherical symmetric fast
wind, and an adiabatic (non-radiative) shock. The shock then expands with a
velocity of 
$${\rm v_{shock} \approx 
\left(\frac{\dot{M}_{fw}v_{fw}v_{sw}}{\dot{M}_{sw}} \right)^{1/2} }
$$
\noindent
where ${\rm \dot{M}_{fw}}$ ${\rm, \dot{M}_{sw}}$, ${\rm v_{fw}}$, and
${\rm v_{sw}}$ are mass-loss rate and 
wind velocity of the fast wind (fw) and the slow wind (sw), respectively.

If the fast wind is isotropic and does not show any variation
in density as a function of polar angle, then the ratio
of the expansion velocity of the polar outflow (83 km~s$^{-1}$) to that 
of the ring (20 km~s$^{-1}$) gives us directly the ratio of RSG mass loss 
in both directions: \.M$_{\rm{SW}}$(0$^\circ$)/\.M$_{\rm{SW}}$(90$^\circ$)
$\approx$ 16:1. This is in reasonable agreement with the ratios of 20:1
and 10:1 computed for the progenitor of SN\,1987\,A by Blondin \& Lundqvist 
(1993) and by Martin \& Arnett (1995), respectively.
Assuming a fast wind velocity of 800 km~s$^{-1}$, a slow wind velocity of
50 km~s$^{-1}$, and a ratio of the slow wind and fast wind mass-loss 
rates along
the stellar equator of 90:1 (6:1 in polar direction) one is able to
reproduce the observed shock velocities in the framework of this
very simplified model.

The center of the inner ring does not coincide with the position of
Sher \#25 (offset 1$''$ to 2$''$). 
This may be due to a movement of Sher \#25 relative to the
surrounding ISM. Martin \& Arnett (1995) pointed out that such a movement
would produce asymmetric polar outflow structures, which in turn might explain
the different appearance of the northeastern and southwestern polar outflow
filaments of Sher \#25.
However, the interpretation of the filaments may be complicated 
by Sher \#25's apparent location at the edge of a wind-blown cavity
excavated by the central cluster of NGC~3603 (Figure \ref{fig1}). 
This cavity has 
a diameter of 2 pc, a dynamical age of 10$^4$ yr (Clayton 1986) and 
may have been created by the onset of the Wolf-Rayet phase of the three
central stars of NGC 3603 (Drissen et al.\ 1995).


\section{Sher \#25 and its relation to NGC 3603}

What evidence do we have that Sher \#25 is indeed a member of the
giant H{\sc ii} region NGC 3603?
Firstly, as discussed above, the systemic velocity of the ring is in good
agreement with the line-of-sight velocities of the cloud cores south
of HD 97950.
Secondly, Sher \#25 is not the only BSG in the NGC 3603 region. Spectroscopy
by Moffat (1983) revealed two other BSG in the vicinity of the cluster
core (cf.\ Figure \ref{fig1}). The locations of Sher \#18 (O6If) and Sher \#23 (O9.5Iab) are also
indicated in our V vs.\ V-I CMD (Figure \ref{fig3}). The apparent lack of BSGs
with similar reddening among the ``field stars'' (thin dots in Figure \ref{fig3})
adds additional weight to the assumption that Sher \#25 is associated with
NGC 3603.

Could Sher \#25 then have been born at the same time as the massive central
stars of the cluster? This would require that Sher \#25 originally had been
at least as massive as these central stars and has gone through a violent
Luminous Blue Variable (LBV) phase with a total mass-loss of more than 50\% of
its initial mass (i.e., \.{M} $\ge$ 25 M$_\odot$) before it became a BSG.
Indeed, kinematical age, expansion velocity, and abundances of Sher \#25's
nebula are comparable to those of the AG Car nebula (cf.\ e.g., Leitherer et
al. 1994). With M$_{\rm bol} \approx$ -9\fm1, Sher \#25's luminosity is in the
range of luminosities observed in other LBVs. Thus, an LBV evolutionary 
scenario for Sher \#25 and its circumstellar surrounding cannot be entirely
excluded.

The simultaneous presence of BSGs and stars of MK type O3V (cf.\ Drissen
et al.\ 1995) requires at least two distinct episodes of star formation 
in NGC 3603 separated by $\approx$ 10 Myr. Moffat (1983) and Melnick et al.\ 
(1989) already suggested that star formation in NGC 3603 might not have been
coeval. The starburst in the dense cluster of NGC 3603 might have been 
initiated by the first generation of massive stars through their interaction 
with a dense cloud core. Subsequently, this cloud core developed into the 
present-day starburst. A similar evolutionary scenario has been suggested by 
Hyland et al.\ (1992) in order to explain the starburst in the 30 Dor region.

\section{Sher \#25 and SN\,1987\,A}

Sher \#25's circumstellar nebula resembles that of SN\,1987\,A in many aspects. 
Both objects have an equatorial ring and bipolar nebulae, and show high 
[N{\sc ii}]/\Ha\ ratios, indicating  an enhanced N abundance.
The N enrichment indicates that the rings and the bipolar 
nebulae consist of mass lost from the progenitor at an earlier evolutionary
stage and swept up by the fast BSG wind.
Surface enrichment with material processed by the CNO cycle typically
occurs at the very end of the RSG phase within the last 10$^4$
yr of RSG evolution.

Yet, differences exist. As shown in Table 1,
the expansion velocities and [N{\sc ii}]/\Ha\ ratios are different.
The [\NII]/\Ha\ ratio of the nebula around Sher \#25 
is lower than the ratio observed in the outer northern ring around
SN\,1987\,A despite the lower N abundance in the LMC. 
Furthermore, Sher \#25 exhibits a higher [N{\sc ii}]/\Ha\ ratio in the
bipolar nebulae than in the ring, while SN\,1987\,A shows an opposite trend.

The similarities between the nebulae seem to suggest that Sher \#25 is at a
similar evolutionary stage as the late progenitor of SN\,1987\,A, Sk$-$69\,202.
However, the differences between the nebulae imply that the evolutionary
history differs. This may be due to the abundance differences between the 
young population in the LMC and in the Milky Way, and to mass differences
between the two stars. 

Sher \#25 appears to have been in a rather stable BSG evolutionary phase
during the past decades covered by photometric measurements. Its photometry 
(Table 2) is quite inhomogeneous owing to the variety of different measurement 
techniques used, crowding problems, and spatial variations in the strength 
of the nebular background emission. The overall amplitude of 
variation in V is less than 0\fm25 within the last 35 yr, and less than
0\fm1 over the last 25 yr. The progenitor of SN\,1987\,A did not stand out
as a variable star, either.


Will Sher \#25 explode like Sk$-$69\,202 in the near future? 
The presence of the ring around Sher \#25, the N enrichment in the outflows, 
and the surface enrichment in metals all suggest that Sher \#25 has passed 
at least once through the RSG phase and is now well within its final BSG 
phase, which may last a few 10$^4$ yr in total depending on the initial 
stellar mass (see Martin \& Arnett 1995). Evolutionary models for massive
stars, however, still suffer from many unsolved problems, such as the amount 
of overshooting, semiconvection, mixing, and mass loss, the choice of
convection criteria, and metallicity effects (see, e.g., Langer \& Maeder
1995).  At present, it is premature to predict whether Sher \#25 will
succeed Sk-69 202 to provide another spectacular supernova in the 
southern sky soon.

%
%
\begin{figure*}[ht]
\figcaption{
\Ha+[N{\sc ii}] image (Danish 1.54 telescope,
exposure time 100 s, seeing 0\farcs9) of NGC 3603.
The wind-blown cavity around the central cluster is visible.
The three blue supergiants in this field are marked.
Sher \#25 is located at the northern edge of the cavity.\label{fig1}}
\end{figure*}

\acknowledgements
WB acknowledges support by the Deutsche For\-schungs\-ge\-mein\-schaft 
(DFG) under grant Yo 5/16-1. EKG and YHC were partially supported 
by the NASA grants STI6122.01-94A, NAGW-4519, and NAG 5-3256. EKG acknowledges 
support by the German Space Agency (DARA) under grant 05 OR 9103 0. 
We thank our referee Laurent Drissen for helpful comments.

\begin{deluxetable}{lcc}
\tablecaption{Physical properties of the circumstellar environment of
Sher~\#25 and Sk$-$69$^\circ$202 / SN~1987\,A. We report the ratio of 
[N{\sc ii}]654.8+658.3/H$\alpha$.\label{tbl-2}}
\footnotesize
\tablehead{
&\colhead{Sher\#25}    & \colhead{Sk--69$^\circ$202 / SN\,1987\,A}}
\startdata
 d$_{\rm (inner) ring}$ & 0.4 pc   & 0.4 pc\tablenotemark{a}\\
 v$_{\rm (inner) ring}$ & 20 km s$^{-1}$   & 10 km s$^{-1}$\tablenotemark{b}\\
 v$_{\rm poles}$ & 83 km s$^{-1}$   & \\
 ([N{\sc ii}]/H$\alpha)_{\rm ring}$ & 0.9--1.2 : 1  & 4.2 : 1\tablenotemark{c}\\
 ([N{\sc ii}]/H$\alpha)_{\rm poles}$ & 2.1 : 1  & 2.5 : 1\tablenotemark{c} \\
 ([N{\sc ii}]/H$\alpha)_{\rm background}$ & 0.15 : 1 & 0.09 : 1\tablenotemark{d} \\

\tablenotetext{a}{Panagia et al.\ 1991, Plait et al.\ 1995}
\tablenotetext{b}{Jakobsen et al.\ 1991}
\tablenotetext{c}{Panagia et al.\ 1996}
\tablenotetext{d}{Chu, unpublished}
\enddata
\end{deluxetable}

\begin{deluxetable}{lcccc}
\footnotesize
\tablecaption{Photometric observations of Sher \#25. \label{tbl-1}}
\tablehead{
\colhead{Reference} & \colhead{date}   & \colhead{V}   & \colhead{B-V} 
& \colhead{U-B}}
\startdata
Sher (1965) & $\approx$1962   & 12\fm07 & +1\fm59 & +0\fm19 \\
\tablenotemark{a} van den Bergh (1978)&         &{\it 12\fm08} 
&{\it +1\fm59} &{\it +0\fm19} \\
Moffat (1974)&$\approx$1972   & 12\fm27 & +1\fm36 & +0\fm10 \\
\tablenotemark{a} van den Bergh (1978)  &         &{\it 12\fm38} 
&{\it +1\fm40} &{\it +0\fm29} \\
van den Bergh (1978)& 1976/77 & 12\fm28 & +1\fm38 & +0\fm25 \\
Melnick et al.\ (1989)& 1985 Feb.& 12\fm20 & +1\fm42 & +0\fm13 \\
Moffat et al.\ (1994)& 1991 Feb.& --- & ---\tablenotemark{b} & --- \\
This paper & 1995 March 2& 12\fm31 & \tablenotemark{c} &   \\
\enddata

\tablenotetext{a}{applying photometric transformations as derived by van
den Bergh (1978)}
\tablenotetext{b}{Using HST/PC1 Moffat et al.\ (1994) measured B=13\fm50.}
\tablenotetext{c}{VRI measurements: V-R=+1\fm03, R-I=+1\fm03}
 
\end{deluxetable}


\begin{thebibliography}{}

\bibitem{} Blondin, J.M., \& Lundqvist, P.\ 1993, ApJ, 405, 337

\bibitem{}
Brandner, W., Dottori, H., Grebel, E.K., et al.\ 1997, A\&A, {\it Stellar 
and non-stellar emission line objects in NGC 3603}, to be submitted

\bibitem{} Chevalier, R.A., \& Imamura, J.N.\ 1983, ApJ, 270, 554

\bibitem{} Clayton, C.\ 1986, MNRAS, 219, 895

\bibitem{} Drissen, L., Moffat, A.F.J., Walborn, N.R., \& Shara, M.M.\ 1995, 
AJ, 110, 2235

\bibitem{} Hyland A.R., Straw S., Jones T.J., Gatley I.\ 1992 MNRAS 257, 391

\bibitem{} Jakobsen, P., Albrecht, R., Barbieri, C., et al.\ 1991, ApJ, 369, L63

\bibitem{} Langer, N., \& Maeder, A. 1995, A\&A, 295, 685

\bibitem{} Leitherer C., Allen R., Altner B.\ et al. 1994 ApJ 428, 292

\bibitem{} Martin, C.L., \& Arnett, D.\ 1995, ApJ, 447, 378

\bibitem{} Melnick, J., Tapia, M., \& Terlevich, R.\ 1989, A\&A, 213, 89

\bibitem{} Moffat, A.F.J.\ 1974, A\&A, 35, 315

\bibitem{} Moffat, A.F.J.\ 1983, A\&A, 124, 273

\bibitem{} Moffat, A.F.J., Drissen L., \& Shara M.M.\ 1994 ApJ 436, 183

\bibitem{} Moore, C.E.\ 1959, {\it A Multiplet Table of Astrophysical
Interest}, United States Department of Commerce, Washington, D.C.

\bibitem{} Panagia, N., Gilmozzi, R., Macchetto, F.\ et al.\ 1991, ApJ, 380, L23

\bibitem{} Panagia, N., Scuderi, S., Gilmozzi, R., et al.\ 1996, ApJ, 459, L17

\bibitem{} Plait, P.C., Lundqvist, P., Chevalier, R.A., \& Kirshner, R.P.\ 1995,
ApJ, 439, 730

\bibitem{} Sahai, R., Trauger, J.T., \& Evans, R.W.\ 1995, BAAS, 27, 1344 

\bibitem{} Sher, D.\ 1965, MNRAS, 129, 237

\bibitem{} van den Bergh, S.\ 1978, A\&A, 63, 275

\end{thebibliography}
\end{document}